\documentclass[journal]{IEEEtran}
\usepackage{microtype}
\usepackage{graphicx}
\usepackage{graphicx}
\usepackage{booktabs} % for professional tables
\usepackage{hyperref}

\usepackage{amsthm}
\usepackage{amsmath}
\usepackage{bbm}
\usepackage{subcaption}
\usepackage{amssymb}
\usepackage{algorithm}
\usepackage{algorithmic}
\usepackage{float}
\usepackage[utf8]{inputenc}
\usepackage[english]{babel}
\usepackage{multicol}
\usepackage{flushend}
\newlength\myindent
\newtheorem{theorem}{Theorem}
\AfterEndEnvironment{theorem}{\noindent\ignorespaces}

\newtheorem{lemma}{Lemma}
\AfterEndEnvironment{lemma}{\noindent\ignorespaces}

\newtheorem{corollary}{Corollary}[theorem]
\AfterEndEnvironment{corollary}{\noindent\ignorespaces}

\newtheorem{definition}{Definition}[section]
\AfterEndEnvironment{definition}{\noindent\ignorespaces}

\usepackage{cite}
\usepackage{hyperref}
\usepackage{tabularx}

\begin{document}

\title{Neural Joint Entropy Estimation}

\author{Yuval Shalev, Amichai Painsky
        and Irad Ben-Gal%

%  \thanks{Y. Shalev, A. Painsky and I. Ben-Gal are with the Laboratory for AI, Machine Learning, Business and Data 
%   Analytics, Department of Industrial Engineering, The Tel-Aviv University, Ramat-Aviv 6997801, Israel.}% <-this % stops a space
  \thanks{Y. Shalev, A. Painsky and I. Ben-Gal are with the Department of Industrial Engineering, Tel-Aviv University, Ramat-Aviv 6997801, Israel. Correspondence to:$<$yuvalshalev@mail.tau.ac.il$>$.}% <-this % stops a space

% \thanks{Manuscript received XXXX; revised XXXX.}
}

\maketitle

\IEEEpeerreviewmaketitle

% \markboth{IEEE TRANSACTIONS ON NEURAL NETWORKS AND LEARNING SYSTEMS,~Vol.~XXX, No.~XXX, XXXXmonth~2020}%

\markboth{}%
{Shalev \MakeLowercase{\textit{et al.}}: Joint Entropy Estimation with Neural Networks}




\begin{abstract}
Estimating the entropy of a discrete random variable is a fundamental problem in information theory and related fields. This problem has many applications in various domains, including machine learning, statistics and data compression. Over the years, a variety of estimation schemes have been suggested. However, despite significant progress, most methods still struggle when the sample is small, compared to the variable's alphabet size. In this work, we introduce a practical solution to this problem, which extends the work of McAllester and Statos (2020). The proposed scheme uses the generalization abilities of cross-entropy estimation in deep neural networks (DNNs) to introduce improved entropy estimation accuracy. Furthermore, we introduce a family of estimators for related information-theoretic measures, such as conditional entropy and mutual information. We show that these estimators are strongly consistent and demonstrate their performance in a variety of use-cases. First, we consider large alphabet entropy estimation. Then, we extend the scope to mutual information estimation. Next, we apply the proposed scheme to conditional mutual information estimation, as we focus on independence testing tasks. Finally, we study a transfer entropy estimation problem. 
The proposed estimators demonstrate improved performance compared to existing methods in all tested setups. 
\end{abstract}


% \fi

\section{Introduction}
\label{intro}

\IEEEPARstart{E}{ntropy} is one of the basic building blocks of information theory \cite{cover2012elements}. It quantifies the minimum average number of bits required to represent an event  that follows a given probability distribution rule. Many important information-theoretic measures such as mutual information (MI) and conditional MI (CMI) include marginal, conditional and joint entropies. These measures have many applications in machine learning, such as feature selection \cite{fleuret2004fast,peng2005feature}, representation learning \cite{chen2016infogan,oord2018representation} and  analyses of the learning mechanism \cite{tishby2000information, tishby2015deep}. 

One of the first and basic entropy estimation methods is the classic plug-in scheme. In this scheme, an empirical distribution replaces the true (unknown) probability rule, and the corresponding empirical entropy is  the estimated entropy. Unfortunately, this estimation scheme suffers from a negative bias 
\cite{paninski2003estimation, wu2016minimax}, leading to limited outcomes. A variety of parametric and nonparametric methods have been  proposed to improve the entropy estimation, such as in \cite{chao2003nonparametric,paninski2003estimation,wu2016minimax}. Recently, a neural network-based method was proposed to estimate entropy by minimizing  the cross-entropy (CE) loss\cite{mcallester2020formal} as an upper bound of the entropy. The CE measures the average number of bits required to represent an event that is generated from a probability distribution $P$ by a different probability distribution $Q$. CE has its minimum when $P=Q$. Thus, minimizing CE implies searching for a $Q$ that is as similar as possible in a log-loss \cite{painsky2018universality,painsky2019bregman} sense to $P$. This approach has several advantages. First, it uses the generalization power of neural networks and their universality \cite{hornik1989multilayer,zhang2016understanding,chong2020closer}. Second, CE is less prone to negative bias and high variance in large entropy values \cite{mcallester2020formal}. However, this approach has certain limitations. 
First, it requires prior assumptions on the true underlying distribution, as  discussed in Section \ref{back}. Second, the statistical properties of this CE estimator are currently unexplored. Therefore, the existence of a neural network-based estimator that can provide an accurate estimation of entropy, is not guaranteed.  

These challenges in entropy estimation are also related to other information-theoretic measures. For example, one of the most common MI estimation schemes is the K-nearest neighbour (KNN)  estimator \cite{kraskov2004estimating}. This estimator was shown to introduce a significant negative bias in setups with high dependencies between the variables, resulting in large MI values \cite{belghazi2018mine}. Neural-network-based approaches have been recently proposed to overcome this problem using variational bound optimization  \cite{belghazi2018mine,poole2019variational,song2019understanding}. Although a significant improvement in the MI estimation has been achieved, the results are not yet satisfying and suffer from theoretical limitations that are primarily manifested in large MI values \cite{mcallester2020formal, poole2019variational}. There is also a large body of work on fundamental estimation bounds for different information-theoretic measures (see \cite{jiao2015minimax,wu2016minimax} and related work).

In this paper, we address the inherent estimation challenges discussed above. The proposed estimation scheme focuses on joint entropy estimation. This problem is similar to the standard entropy estimation problem as any univariate random vector may be represented, for example, as a binary multivariate vector. In particular, we combine the chain rule with the CE loss minimization procedure using neural networks to obtain a more accurate joint entropy estimation. We denote this estimation procedure as the Neural Joint Entropy Estimator (\textit{NJEE}). We study the properties of \textit{NJEE} and show that it is strongly consistent. In a similar manner, we obtain the conditional \textit{NJEE} (\textit{C-NJEE}),  as an estimator for the joint conditional entropy between two or more multivariate variables.

Having these two estimators, it is straightforward to estimate the MI between two random variables. Adding a second conditioning variable results in the CMI estimator. Additionally, we apply the proposed scheme to transfer entropy estimation (TE). Given two time series, the TE is defined as the CMI between the "past" of the first series and the "future" of the second series given its "past". TE is used to explore the information flow and causality among time-dependent data in neuroscience \cite{vicente2011transfer, wollstadt2014efficient}, finance \cite{marschinski2002analysing, dimpfl2014impact}, process control \cite{barnett2009granger, duan2013direct} and many other applications. We show that by using an autoregressive neural network model, such as a recurrent neural network, \textit{C-NJEE} can be used for efficient TE estimation.

The advantages of the estimators proposed in this paper are demonstrated in various use-cases. First, we study the entropy estimation of a discrete random variable with a large alphabet size. Applying \textit{NJEE} to this problem, we outperform existing methods when the sample size is much smaller than the alphabet size. Further, we focus on MI estimation between two multivariate variables. A commonly used toy problem is used for this task. The performance of the proposed MI estimator demonstrates improved results  in terms of lower bias and variance, compared to existing methods. This result is specifically manifested in larger values of MI. Next, we demonstrate the performance of the suggested CMI estimator, as we focus on conditional independence tests. We study a real protein dataset where dependencies among the variables (protein elements) are known. Also, the proposed estimation scheme demonstrates better results than existing methods. Finally, the CMI estimator is applied to a TE estimation task. Specifically, we study a real financial dataset of stock index prices and show that the \textit{C-NJEE}-based estimation provides additional insights on the information flow between the time series that are not discovered by the other methods. These insights are in line with domain knowledge and the world financial timeline.    

To summarize, the contributions of this paper are threefold. First, we extend the work of \cite{mcallester2020formal} and introduce  strongly consistent estimators for joint entropy and conditional joint entropy. The proposed estimators, \textit{NJEE} and \textit{C-NJEE}, are based on minimization of the CE loss while applying the entropy chain rule property.
Second, we use these estimators to obtain estimators for related measures such as MI, CMI and TE. Third, we propose a practical implementation scheme of these estimators that demonstrates better performance than existing methods on various tasks and datasets.

The remainder of this paper is organized as follows. Related works on entropy, MI, CMI and TE estimation are discussed in Section \ref{related work}. In Section \ref{back}, definitions and related mathematical overview are given to support the scheme and ideas proposed in this paper. The primary results  are shown in Section \ref{measuring the joint}. An empirical study of various tasks and comparisons with different benchmark methods are provided in Section \ref{experiments}. We conclude this paper in Section \ref{conclusion}.

\section{Related Work}
\label{related work}
Estimating information-theoretic measures is a well-studied problem. We refer the reader to \cite{paninski2003estimation, kraskov2004estimating, wu2016minimax, verdu2019empirical, bossomaier2016introduction, poole2019variational} for a comprehensive review of these measures. 
The following literature review focuses on estimators that are relevant for this work. 
\subsection{Entropy Estimation in Large Alphabet}
\label{subsec:uni benchmark}
As mentioned in Section \ref{intro}, the simplest method to estimate the entropy of a discrete random variable is the so plug-in estimator \cite{cover2012elements}. The Miller-Madow estimator \cite{miller1955note} adds a bias correction to the plug-in estimator. This correction depends on the ratio between the number of symbols from the alphabet that appear at least once in the sample and the sample size. 

More recently, \cite{chao2003nonparametric} proposed an estimator for the entropy of species in a community (in this biological context, the entropy is called the diversity index), where the number of species (alphabet size) is large and unknown. This estimator is based on the Horvitz-Thompson estimator for population size and the Good-Turing estimator for the probability of unseen events. In \cite{wu2016minimax}, an entropy estimator is obtained using a polynomial approximation for the terms in the entropy sum that involve small probabilities with respect to $\log k$, where $k$ is the alphabet size. For larger probabilities, an unbiased plug-in estimator is used that is similar to the Miller-Madow estimator. Thus, improved results are demonstrated on simulated data of discrete random variables with large alphabet sizes where many symbols have relatively low probability.

\subsection{MI and CMI Estimation}
One of the most popular MI estimators in recent years is the KNN-based KSG estimator \cite{kraskov2004estimating}, which uses  KNN-based density estimation over a shared space of the marginal and conditional entropy. Using the connection between the MI and entropy  (see \ref{definitions}), the entropies' bias terms are subtracted to provide a more accurate MI estimation. This metric suffers from the curse of dimensionality and underestimates the MI when the interaction between the variables is strong \cite{gao2015efficient}.

The recent advances in deep learning motivated various researchers to address the dimensionality problem by estimating
the MI with neural networks. This is usually obtained by finding variational lower bound for the MI (typically, a differentiable
function that its supremum is the MI). These functions are approximated by neural networks to maximize the lower bound \cite{belghazi2018mine, poole2019variational, song2019understanding}. These methods yield improved results compared to the KNN-based estimator, but are quite limited  when estimating large MI values,  since their estimation complexity increases exponentially with the number of samples \cite{mcallester2020formal, poole2019variational}. To overcome this  problem, \cite{mcallester2020formal} proposed using the CE as an upper bound for the entropy and minimize it by training a neural network. Thus, an MI estimate is obtained by subtracting the estimated conditional entropy from the estimated marginal entropy. This approach underlines the proposed estimation scheme as discussed in further detail in subsection \ref{cross_based_mi}. A similar approach for MI estimation using  the softmax function (e.g., as the output layer in a neural network), was suggested in \cite{qin2019rethinking}. However, this scheme is limited to the case where the input variable is multivariate, while the target variable is univariate. Classifier based conditional mutual information (\textit{CCMI}) was proposed in \cite{mukherjee2019ccmi}. A two-sample classifier was used to distinguish between samples from the joint distribution  and samples from the marginal distribution. Combining conditional generative models (e.g., Conditional Generative Adversarial Networks (CGAN) or Conditional Variational Autoencoders (CVAE)), an estimator for the CMI was developed. This approach introduced a significant improvement over other recently proposed methods.

\subsection{TE Estimation}
The TE is defined as a form of CMI between time series. Specifically, $TE(Y_{future};X_{past})  = CMI(Y_{future},X_{past}|Y_{past})$ (see a formal definition of TE in Subsection \ref{definitions}). There are two primary approaches for TE estimation. The first approach considers every variable in every timestamp as a separate variable, and uses any MI or CMI estimator to estimate the TE \cite{runge2012escaping, montalto2014mute, zhang2019itene}. The second approach applies a sequential model that considers the time dependencies among different time lags to extract an estimator for the TE and its related measures \cite{jiao2013universal, shalev2019context}. As a representative of the first approach, a recently proposed estimator \cite{zhang2019itene} applies a neural network two-sample classifier to estimate the TE. Using the second approach, the Context Tree Weighting (CTW) algorithm \cite{willems1998context} is utilized in \cite{jiao2013universal} for directed information  estimation (a closely related measure to TE \cite{liu2012relationship}). Both works investigate a financial time series of index prices to evaluate their estimators. we use the same dataset to evaluate the proposed method. 

\section{Background}
\label{back}
\subsection{Notations}
\label{notations}
The following notations are used throughout this paper. A univariate discrete random variable is denoted by an upper-case letter (e.g., $X$), that obtains  values $x$ from the alphabet $\mathcal{A}_x=\{1,\ldots, a_x\}$. A multivariate variable with dimensions $d_x$ is denoted by an underline, (e.g., $\underline{X}$), where its values are denoted by underlined lower-case letter $\underline{x}$. The $m^{th}$ component of $\underline{X}$ is denoted as $X_m$, which obtains values $x_m$ from the alphabet $\mathcal{A}_{x_m}=\{1\ldots a_{x_m}\}$ which can be different for different values of $m$. The vector of the first $k$ components of $\underline{X}$ is denoted by $\underline{X}^k$. 
 
We denote $\widehat{H}_n(\underline{X})$ as the estimator of  $\underline{X}$'s entropy given a sample $S=\{\ldots\}_{i=1}^n$, where it is implied from the text that $S$ is a collection of $n$ samples of $\underline{X}$. This notation holds for other estimators as well. For example, $\widehat{I}_n(\underline{X};\underline{Y}|\underline{Z})$ is an estimator of the CMI between $\underline{X}$ and $\underline{Y}$ given $\underline{Z}$, from a collection of $n$ samples from the joint distribution of $\underline{X}$, $\underline{Y}$ and $\underline{Z}$. To avoid an overload of notation, we denote $x_i$ as the $i^{th}$ sample in $S$, while $X_m$ is the $m^{th}$ component of the random vector $\underline{X}$.

For the time notation, a multivariate variable in time $t$ is represented by a bracket index, e.g., $\underline{X}_{(t)}$ and a matrix that represents its past $l$ time lags is represented by $X_{(t)}^{(l)}=[\underline{X}_{(t-l)}\ldots\underline{X}_{(t)}]$.

\subsection{Definitions} 
\label{definitions}
Let $\underline{X}$ be a discrete random variable that follows a probability distribution $P(\underline{X})$. Shannon's entropy is defined as: 
\begin{align}
\begin{split}
 &H(\underline{X})  =  -\mathbb{E}_{P(\underline{X})} \left[  \log{P(\underline{x})} \right] \label{H}.
\end{split}
\end{align}
The entropy (\ref{H}) can be represented by the chain-rule 
\begin{align}
\begin{split}
 &H(\underline{X}) = H(X_1, X_2, \ldots, X_{d_x})  = \\ &\sum_{m=1}^{d_x}H(X_m|X_{m-1}, \ldots, X_1),
 \label{H_joint} 
\end{split}
\end{align}
where $H(X_1|X_0)$ abbreviates $H(X_1)$.

The CE between any two distribution functions $P(\underline{X})$ and $Q(\underline{X})$ is defined as:
\begin{align}
& CE(Q(\underline{X})) =  -\mathbb{E}_{P(\underline{X})}\left[\log Q(\underline{X})\right], 
\label{eq:ce}
\end{align}
where the expectation is over the distribution of $\underline{X}$, namely, $P(\underline{X})$.

The following inequality holds for every pair of distributions  $P(\underline{X})$ and $Q(\underline{X})$:
\begin{align}
CE(Q(\underline{X})) \geq H(\underline{X}),
\label{eq:ce greater entropy}
\end{align}
Where an equality is obtained for $Q(\underline{X})=P(\underline{X})$.

A related measure to CE is the Kullback-Leibler divergence ($D_{KL}$) between $P(\underline{X})$ and $Q(\underline{X})$ 
\begin{align}
  D_{KL}(P(\underline{X})||Q(\underline{X}))=\mathbb{E}_{P(\underline{X})}\left[\log \frac{P(\underline{X})}{Q(\underline{X})}\right] .
  \label{KLD}
\end{align}
The $D_{KL}$ is a nonnegative measure and equals zero iff $P(\underline{X})=Q(\underline{X})$.

The MI, denoted as $I(\underline{X}; \underline{Y})$, quantifies in bits the entropy reduction in $\underline{X}$  given the knowledge obtained from another random variable $\underline{Y}$, i.e., 
\begin{align}
\begin{split}
 I(\underline{X}; \underline{Y})  = H(\underline{X}) - H(\underline{X}|\underline{Y}).
\label{eq:MI}
\end{split}
\end{align}

Another important measure that is represented by the difference of entropies is  conditional mutual information (CMI)
\begin{equation}
\begin{aligned}
 I(\underline{X}; \underline{Y}|\underline{Z})  = H(\underline{Y}|\underline{Z}) - H(\underline{Y}|\underline{X}, \underline{Z}). \label{CMI}
\end{aligned}
\end{equation}
CMI is also used to evaluate the TE, which is defined as 
\begin{align}
\begin{split}
   TE_{X\rightarrow Y} = I(\underline{X}_{(t)}^{(k)}; \underline{Y}_{(t+1)}| \underline{Y}_{(t)}^{(l)}).
    \label{cmince_te}
    \end{split}
\end{align}
 Assuming discrete time, the $TE_{X\rightarrow Y}$ is the CMI between the past $k$ time lags of $\underline{X}$ and $\underline{Y}$ at time $t+1$ given the past $l$ time lags of $\underline{Y}$.

\subsection{CE-Based Entropy}
\label{cross_based_mi}
Let $P(\underline{X})$  be the distribution function of $\underline{X}$. Let $T_{\theta}(\underline{X})$ be a neural network that approximates it. In \cite{mcallester2020formal},  the following upper bound for the entropy of $\underline{X}$ was proposed 
\begin{align}
    H_{\Theta}(\underline{X}) = \inf\limits_{\theta\in \Theta}  CE(T_{\theta}(\underline{X})),
    \label{inf_over_theta}
\end{align}
and $H_{\Theta}(\underline{X})=H(\underline{X})$ iff $P(\underline{X})=T_{\theta}(\underline{X})$.
Given a sample $S$ of size $n$, the sample mean is used to estimate the CE, 
\begin{align}
    \widehat{CE}_n(T_{\theta}(\underline{X})) = -\frac{1}{n}\sum\limits_{i=1}^n \log T_{\theta}(\underline{x}_i).
    \label{estimated_cross}
\end{align}
Then, an estimator of the entropy is:
\begin{align}
    \widehat{H}_n(\underline{X}) = \inf\limits_{\theta\in \Theta}\widehat{CE}_n( T_{\theta}(\underline{X})).
    \label{estimated_cross}
\end{align}
In \cite{mcallester2020formal}, the authors suggest an entropy estimator based on the above. However, they require prior knowledge of $P(X)$ for the training of their neural network.

\section{Measuring the joint entropy with neural networks}
\label{measuring the joint}
In this section we discuss the primary concepts of this paper. First, the neural network classifier and its respective CE are formally defined. Then, the neural joint entropy estimator is introduced. Next, we define a strongly consistent estimator and show that the proposed joint entropy estimator satisfies this property. We also provide an algorithmic implementation of the proposed estimator and discuss practical aspects of its implementation. Next,  estimator for the joint conditional entropy is provided with the corresponding algorithmic implementation. Using  the estimators of the joint entropy and the conditional joint entropy, estimators for MI, CMI and TE are obtained.   

\subsection{Neural Network Classifier and Classification CE}
The following basic definitions are used throughout this section. 

\begin{definition}{(Neural network classifier).}
Let $G_\theta(Y|\underline{X})$ be a neural network model with a random variable input $\underline{X}$ and parameters
 $\theta$ in a compact domain $\Theta\in\mathbb{R}^k$.  The outputs of $G_\theta(Y|\underline{X})$ are defined over the probability simplex: $\{G_\theta(y|\underline{x})\in \mathbb{R}^{a_y}:\sum_{y=1}^{a_y} G_\theta(y|\underline{x})=1, G_\theta(y|\underline{x})\geq 0\}$.
 \label{def:classifier}

\end{definition}
Next, we define the CE of this classifier.
\begin{definition}(Classifier CE).
Let $G_\theta(y|\underline{x})$ be a neural network classifier. The CE of this classifier is defined as
\begin{align}
   CE(G_\theta(Y|\underline{X}))= -\mathbb{E}_{P(\underline{X},Y)}\log G_\theta(y|\underline{x}),
    \label{common ce}
\end{align}
where $Y\in\mathcal{A}_y=\{1,\ldots , a_y\}$, $a_y\geq 2$.
\label{def:classification_ce}
\end{definition} 
We assume that $-\log(G_\theta(y|\underline{x}))\leq \eta $ for all $ \underline{x}\in \underline{X}$ and for all $\theta\in\mathbb{R}^k$,  for any value of $Y$. Practically, this assumption is used in many model training procedures to avoid an unbounded loss \cite{painsky2019bregman}.
The empirical estimator of this CE is given by \cite{zhang2018generalized}, namely
\begin{align}
    \begin{split}
        &\widehat{CE}_n( G_{\theta}(Y|\underline{X})) =  -\frac{1}{n}\sum\limits_{i=1}^n \log(G_{\theta}(y_i|\underline{x}_i)).
        \label{estimated_cross_multi}
    \end{split}
\end{align}
 
 \subsection{Neural Joint Entropy Estimation}
 Using (\ref{H_joint}) and  Definitions  \ref{def:classifier} and \ref{def:classification_ce}, we define the estimator of the joint entropy.
\begin{definition}(Neural Joint Entropy Estimator (\textit{NJEE})). Let $\widehat{H}_n(X_1)$ be an estimated marginal entropy of the first components  in $\underline{X}$ and let $G_{\theta_m}(X_m|\underline{X}^{m-1})$ be a neural network classifier. Then, \textit{NJEE} is defined as
\begin{align}
        \begin{split}
           \widehat{H}_n(\underline{X})=  \widehat{H}_n(X_1)+\sum\limits_{m=2}^{d_x}\widehat{CE}_n( G_{\theta_m}(X_m|\underline{X}^{m-1}).
        \end{split}
        \label{def:NJEE}
    \end{align}
\end{definition}
In words, the joint entropy estimator consists of a marginal estimator for the first component, followed by estimators for the conditional entropies $H(X_m|X^{m-1})$, for $m=2,\ldots , d_x$.

\begin{definition}
 (Strong consistency (following \cite{belghazi2018mine})). The estimator
$\widehat{H}_n(\underline{X})$
is strongly consistent if for all $\epsilon, \delta>0$ and a constant $C>0$, there exists
a positive integer $N$ and a choice of a neural network such
that:\[\forall n \geq N,  |H(\underline{X}) - \widehat{H}_n(\underline{X})|\leq\ C\cdot\epsilon + \delta, a.e.\]
\end{definition}

\begin{theorem}
\textit{NJEE} is strongly consistent.
\label{theorem:join strong consistency}
\end{theorem}

\subsection{Proof of Strong Consistency Property}
In this section we follow the scheme  shown in \cite{belghazi2018mine} to prove Theorem \ref{theorem:join strong consistency}. This proof  includes the following main steps:

\begin{enumerate}
    \item Connecting the true CE of a classifier-based neural network and the conditional entropy $H(Y|\underline{X})$ (Lemmas \ref{lemma:kl P and G} and \ref{lemma:ce to H}).
    \item Showing the convergence of the empirical CE to the true CE (Lemma \ref{lemma:estimate}).
    \item Showing that the empirical CE can approximate with high accuracy the  conditional entropy (Lemma \ref{lemma: univariate cond entropy}).
    \item Applying the chain-rule property and the previous steps to show that the proposed estimator of the joint entropy is strongly consistent.
\end{enumerate}

We begin with the first step. Formally, since neural networks are universal approximation functions \cite{hornik1989multilayer, zhang2016understanding, chong2020closer}, the following  holds:
\begin{lemma}
\label{lemma:kl P and G}
\leavevmode
For any $\epsilon > 0$, and any conditional distribution function $P(Y|\underline{X})$, there exists a neural network $G_\theta(Y|\underline{X})$ such that:
\begin{align}
  D_{KL}(P(Y|\underline{X})||G_\theta(Y|\underline{X}))\leq\frac{\epsilon}{2}, a.e.   
\end{align}

\end{lemma}
That is, it is possible to find a neural network that can approximates $P(Y|\underline{X})$ in any desired approximation level.

The next Lemma states that the CE can be used to estimate the conditional entropy. 
\begin{lemma}
\label{lemma:ce to H}
\leavevmode
Let $P(Y|\underline{X})$ be a conditional distribution and let $H(Y|\underline{X})$ be the entropy associated with this distribution. Then, for any $\epsilon > 0$, there exists a neural network $G_\theta(Y|\underline{X})$ such that 
\begin{align}
  |CE\left(G_\theta(Y|\underline{X})\right) - H(Y|\underline{X})| \leq\frac{\epsilon}{2}, a.e.  
\end{align}

\end{lemma}
The proof of this lemma follows the ideas shown in
 \cite{mcallester2020formal}).
\begin{align}
\begin{split}
&H(Y|\underline{X}) =  \mathbb{E}_{P(\underline{X}, Y)}\log \frac{1}{P(y|\underline{x})} = \\ & \mathbb{E}_{P(\underline{X}, Y)}\log \frac{1}{G_\theta(y|\underline{x})}\frac{G_\theta(y|\underline{x})}{P(y|\underline{x})} = \\
&\mathbb{E}_{P(\underline{X}, Y)}\log \frac{1}{G_\theta(y|\underline{x})} - D_{KL}(P(y|\underline{x})||G_\theta(y|\underline{x})) \geq 
 \\
&CE(G_\theta(Y|\underline{X})) - \frac{\epsilon}{2},
\end{split}    
\end{align}
 where the last line follows Lemma \ref{lemma:kl P and G}. As shown in (\ref{eq:ce greater entropy}), we have that 
\begin{align}
    CE(G_\theta(Y|\underline{X})) - H(Y|\underline{X})\geq 0 ,
\end{align}
 therefore \[|CE\left(G_\theta(Y|\underline{X})\right) - H(Y|\underline{X})| \leq\frac{\epsilon}{2}.\] 

The empirical estimator for this classifier CE is obtained from (\ref{estimated_cross_multi}).  The conditions for the convergence of this estimator are defined by the uniform law of large numbers. 
\begin{lemma}
\label{lemma:uniform}
\leavevmode
The uniform law of large numbers \cite{newey1994large}. 
 Let $\Theta$ be a compact set of parameters. Let $f_\theta(\underline{x}_i)$ be a continuous function at each $\theta\in\Theta$ and $\underline{x}_i\in\underline{X}$ . Assume there exists an upper bound $\eta(\underline{X})$ such that $\|f(\underline{x})\|\leq \eta(\underline{x})$ for all $\theta\in\Theta$ and $\mathbb{E}[\eta(\underline{X})]<\infty$. Then, $E[f_\theta(\underline{X})]$ is continuous and
\begin{align}
 \sup_{\theta\in\Theta}\|\frac{1}{n}\sum\limits_{i=1}^{n}f_\theta(\underline{x}_i) - \mathbb{E}[f_{\theta}(\underline{X})]\|\overset{p}{\to} 0.   
\end{align}
\end{lemma}
Using Lemma \ref{lemma:uniform}, the convergence of the classifier CE is obtained  
\begin{lemma}
\label{lemma:estimate}
\leavevmode
For any $\epsilon > 0$ and $\forall\theta\in\Theta$, there exists a positive integer $n\geq N$ such that:
\begin{align}
  P(|\widehat{CE}_n( G_\theta(Y|\underline{X})) - CE(G_\theta(Y|\underline{X}) ) |\leq\frac{\epsilon}{2})=1.  
\end{align}
\end{lemma}
The proof of this Lemma is an immediate application of (\ref{estimated_cross_multi}) with 
\begin{align}
 f_\theta((\underline{x}_i, y_i)) = -\log(G_{\theta}(y_i|\underline{x}_i)).  
\end{align}
since $-\log(G_\theta(y_i|\underline{x}_i))\leq \eta$, then $f_\theta((\underline{x}_i, y_i))\leq \eta$  and Lemma \ref{lemma:uniform} holds.
\begin{lemma}
\label{lemma: univariate cond entropy}
The estimator $\widehat{CE}_n( G_\theta(Y|\underline{X}))$ is strongly consistent. That is,  for all $\epsilon > 0$, there exists a positive integer $n\geq N$ and a choice of neural network such that:
\begin{align}
  |H(Y|\underline{X}) - \widehat{CE}_n( G_\theta(Y|\underline{X}))| \leq \epsilon,a.e.  
\end{align}
\end{lemma}

This lemma is obtained using the triangular inequality with Lemmas \ref{lemma:ce to H} and \ref{lemma:estimate}:
\begin{align}
\begin{split}
&|H(Y|\underline{X}) - \widehat{CE}_n( G_\theta(Y|\underline{X}))|\leq \\
&|CE\left(G_\theta(Y|\underline{X})\right) - H(Y|\underline{X})| + \\
&|\widehat{CE}( G_\theta(Y|\underline{X})) - CE(G_\theta(Y|\underline{X}) ) |\leq\epsilon.
\end{split}    
\end{align}
Restating (\ref{H_joint}),
\begin{align}
  H(\underline{X})   = H(X_1) +     \sum\limits_{m=2}^{d_x}H(X_m|\underline{X}^{m-1}).
  \label{eq:joint_x_m}
\end{align}
Suppose there exists $d_x-1$ neural networks that approximate each term in the sum with an $\epsilon$ accuracy. Then, the total error of the sum expression is $\epsilon\cdot(d_x-1)$. The marginal entropy $H(X_1)$ is estimated with an estimator $\widehat{H}_n(X_1)$ that guarantees an error that is not larger than certain $\delta>0$. Several estimators can provide such a guarantee, e.g., \cite{paninski2003estimation, wu2016minimax}.   
In this case: 
\begin{align}
\begin{split}
    &|H(\underline{X}) -  \widehat{H}_n(\underline{X})| = 
    |H(X_1) - \widehat{H}_n(X_1) + \\ &\sum\limits_{m=2}^{d_x}H(X_m|\underline{X}^{m-1}) - \sum\limits_{m=2}^{d_x}\widehat{CE}_n( G_{\theta_m}(X_m|\underline{X}^{m-1})|\\
    & \leq  |H(X_1) - \widehat{H}_n(X_1)| + \\ & |\sum\limits_{m=2}^{d_x}H(X_m|\underline{X}^{m-1}) - \sum\limits_{m=2}^{d_x}\widehat{CE}_n( G_{\theta_m}(X_m|\underline{X}^{m-1})| \\&
    \leq \delta + C\cdot\epsilon,
    \end{split}
\end{align}
where $C=d_x-1$. $\square$

\subsection{Algorithmic Implementation of NJEE}
 The implementation of the \textit{NJEE} estimator is described in Algorithm \ref{alg:mince}.
\begin{algorithm}[H]
\caption{NJEE}
\label{alg:mince}
\begin{algorithmic}[1]
\STATE \textbf{input:} Sample $S=\{\underline{x}_i\}_{i=1}^{n}$ from $P(\underline{X})$
\STATE $h_m\leftarrow 0$, \textbf{for} $m=\{1, \ldots, d_x\}$   
\STATE $h_1\leftarrow \widehat{H}_n(X_1)$
\STATE Initialize $\{\theta_{m}\}_{m=2}^{d_x}$
\FOR{$m$ in $2$ to $d_x$}
        \STATE $h_m\leftarrow$ Minimize $\widehat{CE}_n( G_{\theta_m}(X_m|\underline{X}^{m-1}))$
\ENDFOR
\STATE $\widehat{H}_n(\underline{X}) \leftarrow h_1 + \sum\limits_{m=2}^{d_x}h_m$ 
\STATE \textbf{return:} $\widehat{H}_n(\underline{X})$
\end{algorithmic}
\end{algorithm}
Practically, Algorithm \ref{alg:mince} can be implemented in parallel per each value of $m$. Another approach is to use a recurrent neural network (RNN) that replaces the $d_x-1$ networks. In this case, the sequential input to the RNN is the components vector of $\underline{X}$ (e.g., see distribution estimation with RNN in \cite{uria2016neural}). Then, the estimated entropy would be the sum of all the CE losses in every time step. The empirical results of this implementation demonstrate similar performance to Algorithm \ref{alg:mince}.

We also note that by using the CE loss, it is possible to replace the neural network model with any other classifier to estimate the entropy. However, in this case, Lemma \ref{lemma:kl P and G} may not apply, and strong consistency is not guaranteed.

\subsection{Conditional-Neural Joint Entropy Estimation}
The conditional entropy of two multivariate random variables $\underline{X}$ and $\underline{Y}$ is
\begin{align}
  H(\underline{X}|\underline{Y}) =     \sum\limits_{m=1}^{d_x}H(X_m|\underline{Y}, \underline{X}^{m-1}).  
  \label{eq:mult cond entropy}
\end{align}
To estimate (\ref{eq:mult cond entropy}), a slight change is made to \textit{NJEE}, where all components in the proposed estimator are neural networks.
\begin{definition}(Conditional Neural Joint Entropy Estimator (C-NJEE)). Let  $G_{\theta_m}(X_m|\underline{Y},\underline{X}^{m-1})$ be a neural network classifier with inputs $\underline{Y}$ and $\underline{X}^{m-1}$. Then \textit{C-NJEE} is defined as,
\begin{align}
\begin{split}
\widehat{H}_n(\underline{X}|\underline{Y})=\sum\limits_{m=1}^{d_x}\widehat{CE}_n( G_{\theta_m}(X_m|\underline{Y}, \underline{X}^{m-1})).
\end{split}
\label{conditional estimation}
\end{align}
\end{definition}

\begin{corollary}
\leavevmode
\label{corollary: multi cond entropy}
C-NJEE is strongly consistent.
\begin{align}
\begin{split}
&|H(\underline{X}|\underline{Y}) - \sum\limits_{m=1}^{d_x}\widehat{CE}_n( G_{\theta_m}(X_m|\underline{Y}, \underline{X}^{m-1}))|\leq d_x\cdot\epsilon, a.e.
\end{split}
\label{eq: multi cond entropy}
\end{align}

\end{corollary}
The proof of Corollary \ref{corollary: multi cond entropy} is straightforward. Notice that every conditional entropy in the sum expression of (\ref{eq:mult cond entropy}) can be estimated by a classifier CE with $\epsilon$ estimation error. Since there are $d_x$ conditional entropies estimators, the total estimation error of $\widehat{H}(\underline{X}|\underline{Y})$ is $d_x\cdot\epsilon$. The implementation of \textit{C-NJEE} is described in Algorithm \ref{alg:c-njee}.
 \begin{algorithm}[h!]
\caption{C-NJEE}
\label{alg:c-njee}
\begin{algorithmic}[1]
\STATE \textbf{input:} Sample  $S=\{\underline{x}_i, \underline{y}_i\}_{i=1}^{n}$ from $P(\underline{X}, \underline{Y})$ 
\STATE $h_m\leftarrow 0$, \textbf{for} $m=\{1, \ldots, d_x\}$ 
\STATE Initialize $\{\theta_{m}\}_{m=1}^{d_x}$
\FOR{$m$ in $1$ to $d_x$}
        \STATE $h_m\leftarrow$ Minimize $\widehat{CE}_n(G_{\theta_m}(X_m|\underline{Y}, \underline{X}^{m-1}))$
\ENDFOR
\STATE $\widehat{H}_n(\underline{X}|\underline{Y}) \leftarrow  \sum\limits_{m=1}^{d_x}h_m$ 
\STATE \textbf{return:} $\widehat{H}_n(\underline{X}|\underline{Y})$
\end{algorithmic}
\end{algorithm}
  
 We now apply \textit{NJEE} and \textit{C-NJEE} to introduce an estimator for the MI.
\begin{align}
\begin{split}
&\widehat{I}_n(\underline{X}; \underline{Y})=
\widehat{H}_n(X_1) +   \sum\limits_{m=2}^{d_x}\widehat{CE}_n( G_{\theta_m}(X_m|\underline{X}^{m-1}) \\ &- 
\sum\limits_{m=1}^{d_x}\widehat{CE}_n(G_{\theta_m}(X_m|\underline{Y}, \underline{X}^{m-1})),
\end{split}
\label{eq:mi estimation}
\end{align}
Similarly, given a variable $\underline{Z}$, an estimator for the CMI (\ref{CMI}) can be obtained 
\begin{align}
\begin{split}
&\widehat{I}_{n}( \underline{X}; \underline{Y}|\underline{Z})= 
\sum\limits_{m=1}^{d_x}\widehat{CE}_n( 
G(X_m|\underline{Z}, \underline{X}^{m-1})) \\& 
-\sum\limits_{m=1}^{d_x}\widehat{CE}_n( 
G(X_m|\underline{Z}, \underline{Y}, \underline{X}^{m-1})).
\end{split}
\label{eq:mi estimation}
\end{align}
Again, since all models are trained independently, the worst case error of these estimators is the sum of the errors of \textit{NJEE} and \textit{C-NJEE}, thus these estimators are also strongly consistent.

\section{Experiments}
\label{experiments}
In this section we demonstrate the performance of the proposed estimators in various estimation  tasks.

% \footnote{A code implementation of these experiments using \textit{NJEE} and \textit{C-NJEE} is available at \href{https://github.com/YuvalShalev/NJEE.git}{https://github.com/YuvalShalev/NJEE.git}}. 

To apply these estimators, we train a set of neural networks. Unless stated otherwise, the following basic network structure is considered throughout these experiments: An input layer, two fully connected layers with 50 nodes, a ReLU activation function and an output softmax layer. The loss is optimized with the ADAM \cite{kingma2014adam} optimizer with the following parameters $(lr=0.001, \beta_1=0.9, \beta_2=0.999)$. 
\subsection{Entropy Estimation with Large Alphabet}
We begin this experimental section with large alphabet entropy estimation using \textit{NJEE}. Prior to applying \textit{NJEE}, we change the univariate representation values of the  alphabet to their binary representation. Any other small alphabet representation, such as ternary, is also valid. The evaluation is preformed on six simulated studies, most of which were used in previous works (e.g.,  \cite{wu2016minimax}):
\begin{itemize}
    \item Uniform distribution.
    \item Zipf's law distribution with parameters $\alpha=1,2$.
    \item Geometric distribution with $p=1/10^5$.
    \item Symmetric mixture of a Zipf's law distribution ($\alpha=1$) and Geometric distribution ($p=2/10^5$). 
    \item Discrete Laplace (DL), where $DL(X, \sigma)\propto\frac{1}{2\sigma}e^{-\frac{X}{\sigma}}$ and $\sigma=10^{-4}$.
\end{itemize}
The alphabet size of $X$ is set to $10^5$ (excluding the last experiment where the alphabet is not limited). Every simulated study (defined by a distribution type and a sample size) is repeated 100 times. 

\begin{figure*}[t]
\centering
\includegraphics[scale=0.5]{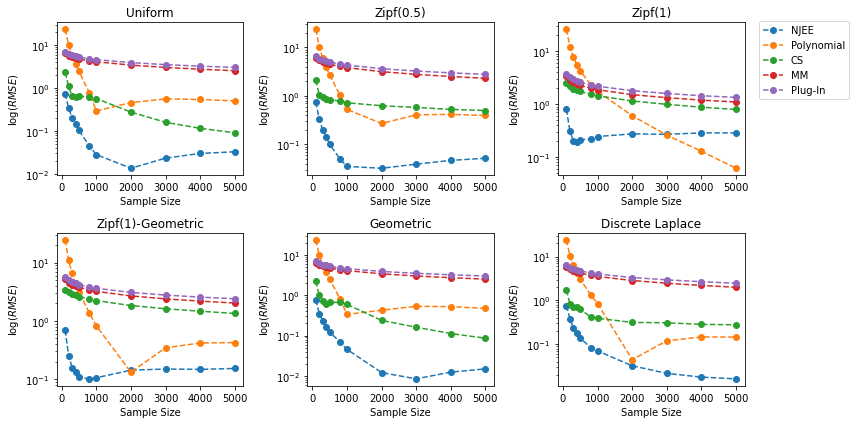}
\caption{The $\log$ of the RMSE of entropy estimations  versus the $\log$ of the sample size for \textit{NJEE} and benchmark methods in different simulated  studies. The results are the average of 100 measurements per each sample size and distribution type.}   
\label{fig:uni_entropy}
\end{figure*}

Figure \ref{fig:uni_entropy} demonstrates the root mean squared (RMSE) of the entropy estimation as a function of the sample size for \textit{NJEE} and other entropy estimators described in Section \ref{subsec:uni benchmark}\footnote{The code of the polynomial method is provided by \cite{wu2016minimax} in \href{https://github.com/Albuso0/entropy}{\textit{https://github.com/Albuso0/entropy}}. See the Entropy R package in  \cite{JMLR:v10:hausser09a} for the implementation of the other benchmark methods.\label{foot_1}}. As shown, \textit{NJEE} demonstrates the lowest RMSE in most cases. Specifically, \textit{NJEE} demonstrates the lowest error in  all the experiments where $n\leq1000$. 

\subsection{Multivariate MI Estimation}
In the following set of experiments we apply the proposed scheme to a simple and commonly used  multivariate MI estimation problems, as used in \cite{belghazi2018mine, poole2019variational, mukherjee2019ccmi}. 
The setup  is defined as follows. Let $\underline{X}$ and $\underline{Y}$ be two random vectors  in $\mathbb{R}^d$ such that
\begin{align}
\begin{split}
\label{mv_normal}
&
\begin{bmatrix}
\underline{X}& \underline{Y}\\
\end{bmatrix}^T \sim \mathcal{N}(0, \Sigma_{XY})\\
&\Sigma_{XY} = 
\begin{bmatrix}
I_d  & \rho I_d\\
\rho I_d & I_d
\end{bmatrix}.
\end{split}
\end{align}
Notice that the correlation between the pairs $(X_{i}, Y_{j})$ is $\rho$ when $i=j$ and zero otherwise. Further,  $Cov(\underline{X})=Cov(\underline{Y})=I_d$, and the MI between $\underline{X}$ and $\underline{Y}$ is thus simply:
\[I(\underline{X}; \underline{Y})=-\frac{d}{2}\cdot\log(1-\rho^2).\] 
In this study, samples are generated from the model above, using different values of
$\rho$ (or equivalently, different values of MI). Since the proposed algorithm is designed for discrete variables, we quantize the samples using a simple binning scheme. Binning continuous data for MI estimation has been extensively studied over the years. The interested reader is referred to \cite{hlavavckova2007causality,dimpfl2013using, dimpfl2014impact,jiao2013universal,montalto2014mute, mcallester2020formal}  for a thorough discussion. 

\begin{figure}[]
\centering
\includegraphics[scale=0.55]{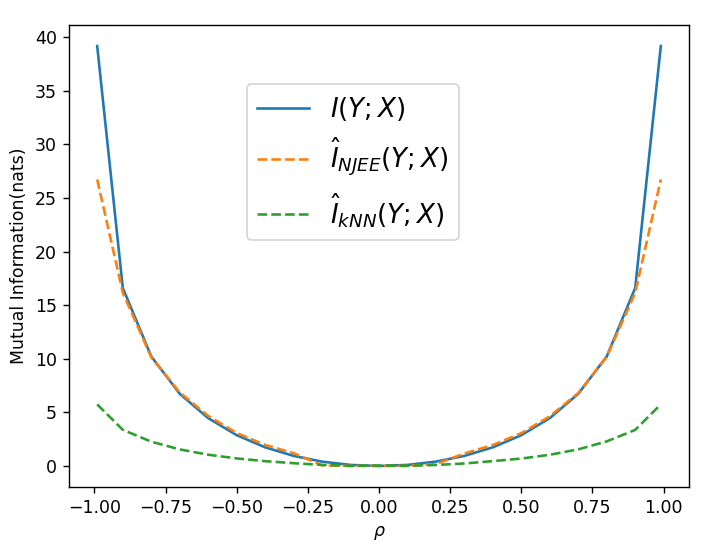}
\caption{MI estimation of the study in (\ref{mv_normal}) with various values of $\rho$. $\widehat{I}_{n}(\underline{X}; \underline{Y})$ is compared to the KNN ($k=3$) method \cite{kraskov2004estimating}. The dimensions of \underline{X} and \underline{Y} are 20.}
\label{fig:njee_knn_mine}
\end{figure}

In Figure \ref{fig:njee_knn_mine}, the \textit{NJEE}-based algorithms are compared to the KNN MI estimation method \cite{kraskov2004estimating}.  With low absolute values of $\rho$, the two methods yield accurate results. As $\rho$ increases (and thus the MI increases), the KNN estimator significantly deviates from the true value, as demonstrated in  \cite{belghazi2018mine}. \textit{NJEE} yields better results for greater MI, similar to \cite{mcallester2020formal}, yet without a prior assumption on the characteristics of the underlined distribution.

Let us now turn to an additional synthetic experiment, following \cite{poole2019variational}. Again, we draw samples from the model described in (\ref{mv_normal}). In this experiment,  we begin with $\rho=0$ and draw a total of 4000 batches with 64 samples in each batch. Then, we estimate the MI from the drawn samples. We increase $\rho$ and repeat the previous step. We terminate at $\rho=1$.
$\widehat{I}_{n}(\underline{X}; \underline{Y})$ is compared to the recently proposed variational methods \footnote{We thank the authors of \cite{poole2019variational} for providing us with the implementation code for the variational methods.}. 
\begin{figure*}[]
\centering
\includegraphics[scale=0.5]{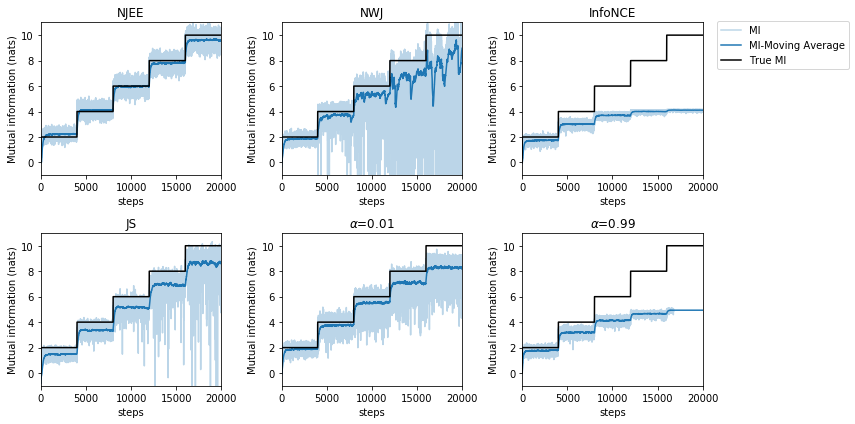}
\caption{MI estimation with \textit{NJEE} versus recently proposed variational methods from \cite{poole2019variational}. Samples from two multivariate random variables in $d=20$ are generated according to (\ref{mv_normal}) with an increasing $\rho$ every 4000 batches. The estimated MI in every batch appears in light blue, the moving average of the MI over a rolling window of 200 batches is shown in dark blue and the true MI value is represented by the black line. The variational bounds shown in this figure are further discussed in the literature (see  \textit{NWJ}\cite{nguyen2010estimating} , \textit{InfoNCE} \cite{oord2018representation}, Jensen-Shannon lower bound (\textit{JS}), and the interpolated bound between \textit{NWJ} and \textit{NCE} with $\alpha=0.01$ and  $\alpha=0.99$ \cite{poole2019variational}).}   
\label{fig:total_mi}
\end{figure*}
As demonstrated in Figure \ref{fig:total_mi}, the results achieved by the proposed estimator exhibits lower bias and variance with respect to the variational benchmark methods. The upper rows of Table \ref{benchmark} demonstrate the best estimation results for each method  obtained by hyperparameter grid search. The proposed \textit{NJEE} scheme yields better results for most MI values ranging from $2$ to $20$. The reasons for the bias and variance errors in the variational bound methods are discussed in \cite{poole2019variational}.

Let us now study estimator sensitivity to invertible transformation, in which we do not expect any change in the MI under such transformations. The cubic transformation $y\Rightarrow z=(Wy)^{3}$ is chosen for this experiment, where $W$ is an invertible $d\times d$ matrix with the entries $w_{ij} \sim \mathcal{N}(0, 1)$. The lower rows of Table \ref{benchmark} summarize the results. As shown, the proposed MI estimator yields identical results to the original problem, while the alternative methods yield lower estimates. Due to stability issues in the benchmark methods, we could not obtain  estimates for the cubic transformation when the underlying MI equals $20.0$ nats. 

\begin{table}[h]
\centering
\caption{Best results of every estimator following a hyperparameter grid search for the Gaussian setup (\ref{mv_normal}) (upper rows) and its cubic transformation (lower rows). The true MI values are shown in the first row. The results of the benchmark methods for 2 to 10 nats are also reported in \cite{poole2019variational}.}
\vskip 0.15in
%\begin{center}
\begin{small}
\label{benchmark}
\begin{sc}
\begin{tabular}{l|cccccr}
\toprule
\multicolumn{7}{r}{True mutual information}\\
&2.0 & 4.0 & 6.0 & 8.0 & 10.0 & 20.0 \\
\midrule
\underline{Gaussian setup}\\
  \textit{NJEE}  & 2.2 & \textbf{4.1} & \textbf{5.9} & \textbf{7.8} & \textbf{9.6} & \textbf{17.8}\\
  $\alpha$  & \textbf{1.9} & 3.8 & 5.7 & 7.4 & 8.8 & 11.7\\
  \textit{JS}  & 1.2 & 3.0 & 4.8 & 6.5 & 8.1 & 15.5\\
  \textit{NWJ}  & 1.6 & 3.5 & 5.2 & 6.7 & 8 & 10.8\\
  \textit{InfoNCE}  & \textbf{1.9} & 3.6 & 4.9 & 5.7 & 6 & 6.2\\
\midrule
\underline{Cubic setup}\\
  \textit{NJEE}  & \textbf{2.2} & \textbf{4.1} & \textbf{5.9} &\textbf{ 7.8} & \textbf{9.6} & \textbf{17.8}\\
  $\alpha$  & 1.7 & 3.6 & 5.4 & 6.9 & 8.2 & - \\
  \textit{JS}  & 1 & 2.8 & 4.5 & 6.1 & 7.6 & -\\
  \textit{NWJ}  & 1.5 & 3.2 & 4.7 & 5.9 & 6.9 & -\\
  \textit{InfoNCE}  & 1.7 & 3.2 & 4.1 & 4.6 & 4.8 & -\\
\bottomrule
\end{tabular}
\end{sc}
\end{small}
%\end{center}
\vskip -0.1in
\end{table}

\subsection{Conditional Independence Testing}
We now investigate the proposed method in conditional independent testing (CIT). CIT is a basic task in statistics with applications to a variety of domains, such as Bayesian networks and causality analysis \cite{ campos2006scoring,sen2017model, zhang2012kernel}. 
In this experiment, we use a flow-cytometry dataset \cite{sachs2005causal}. This dataset describes the connections between eleven proteins in different experimental setups. Sachs et al.,\cite{sachs2005causal}  introduced a consensus Bayesian network (see Figure 3 in their work) that is considered the ground truth of the connections mapping among the proteins.
The flow-cytometry dataset was extensively studied in several works. In \cite{mukherjee2019ccmi}, the authors introduced a CIT method that incorporates a two-sampled classifier and generative models. In \cite{sen2017model}, a  KNN bootstrap and binary classifier procedure was proposed to perform the CIT.  

\begin{figure}[]\centering\includegraphics[scale=0.55]{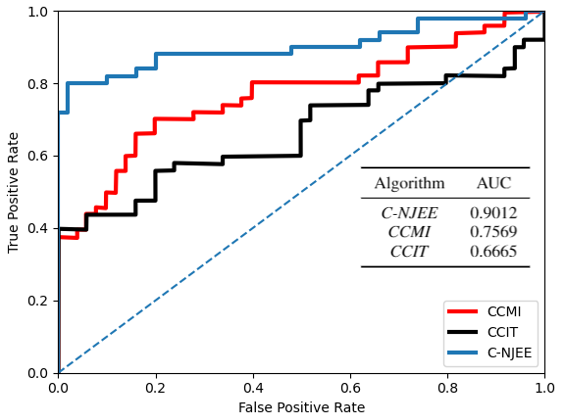}\caption{The ROC curve and the AUC values of \textit{C-NJEE} based estimation, \textit{CCIT} \cite{sen2017model} and \textit{CCMI} \cite{mukherjee2019ccmi} for conditional independence testing task on the flow-cytometry dataset. The dashed line denotes a random model.}\label{fig:cmi}\end{figure}

Before we describe the results of the experiment, we provide some preliminaries on Bayesian networks that are  used for this experiment.  In a Bayesian network, features  are represented by nodes, and their dependencies are represented by edges \cite{ben2008bayesian}. Node A is a parent of node B if there is a directed edge from A to B, and B is considered a child of A. $Y$ is conditionally independent of $\underline{X}$ when $\underline{Z}$ is a subset of the features that holds all available information about $Y$. These features are the parents of $Y$, its children and  the parents of its children (Markov Blanket \cite{statnikov2013algorithms}). Based on these notations, one can choose multiple combinations of dependent and conditionally independent triplet sets of variables. Following the procedures proposed in \cite{sen2017model} and \cite{mukherjee2019ccmi}, 50 dependent and 50 conditionally independent triplets $(\underline{X}, Y, \underline{Z}$) are randomly chosen and their CMI is estimated using $\widehat{I}_{n}(\underline{X}; Y|\underline{Z})$. For every triplet we have the ground truth (dependent/independent), and its corresponding estimate $\hat{I}_n(\underline{X};Y|\underline{Z})$. Since the estimates  $\hat{I}_n(\underline{X};Y|\underline{Z})$ are continuous (nonnegative) numbers, we may set a decision threshold. Specifically, we say that a triplet is conditionally independent if its  $\hat{I}_n(\underline{X};Y|\underline{Z})$ value is lower than a decision threshold $\epsilon$ (and vice versa). Thus, one could construct an ROC curve where every point in the curve represents a value of the threshold $\epsilon$, the value of the false positive rate (the horizontal axis) and the true positive rate (the vertical axis). Figure  \ref{fig:cmi} illustrates the ROC curve and the area under the curve (AUC) values of the independence test performed with  $\widehat{I}_{n}(\underline{X}; Y|\underline{Z})$ and with the benchmarks as reported in \cite{mukherjee2019ccmi}. As shown, $\widehat{I}_{n}(\underline{X}; Y|\underline{Z})$ outperforms the alternative methods. 

\subsection{Estimating TE on Financial Dataset}
Finally, we apply \textit{C-NJEE} to TE estimation.
For this experiment, we study a financial dataset that contains the daily closing prices of the Dow-Jones Index (DJI - the stock index of 30 large companies in the U.S. stock exchange) and the Hang Seng Index (HSI - the stock index of 50 large companies in the Hong-Kong stock exchange)  between  1990 and 2011. As the DJI index is considered more influential than the HSI on the world's financial markets, we expect the transfer entropy  $TE_{DJI\rightarrow HSI}$ to be significantly greater than $ TE_{HSI\rightarrow DJI}$. Additionally, we expect to see changes in the TE that are coordinated with related economic events (e.g., significant financial crises). 

To estimate the TE, we reproduce the preprocessing used in  \cite{jiao2013universal} and \cite{zhang2019itene}, and bin the data to three levels of daily price change. A negative change of more than $-0.8\%$ is denoted by -1,  an absolute change that is below $0.8\%$ is denoted by $0$, and a change that is greater than $0.8\%$ is denoted by $+1$. Then, the \textit{C-NJEE} algorithm is applied with a recurrent neural network that has the following structure: an input layer, followed by an LSTM cell \cite{gers1999learning} with 50 nodes, a fully connected layer with 50 nodes with ReLU activation and an output softmax layer. Input data are divided into sequences of length five (i.e., five consecutive trading days). The optimization procedure includes an ADAM optimizer \cite{kingma2014adam}, with the following parameters:  $lr=0.001, \beta_1=0.9, \beta_2=0.999$.

\begin{figure}[]\centering\includegraphics[scale=0.7]{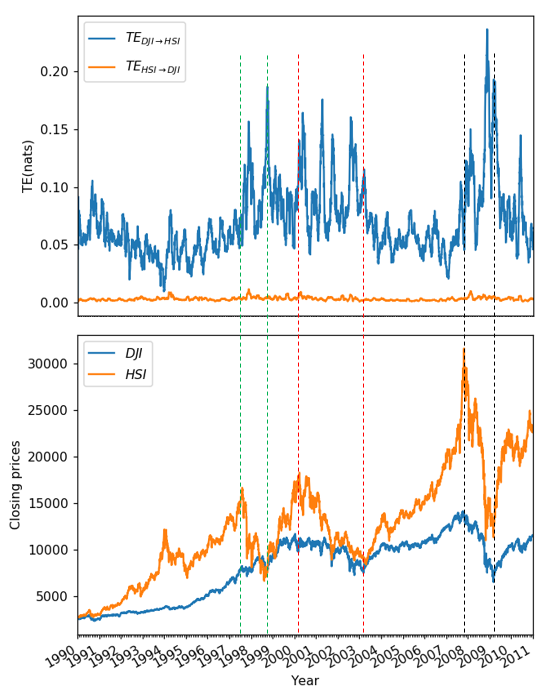}\caption{TE and daily closing prices of the Dow Jones Index (DJI) and the Hang Seng Index (HSI). The upper chart demonstrates the 30-day moving average of the TE estimated by the \textit{C-NJEE} of DJI to HSI ($DIJ\rightarrow HSI$) and in the opposite direction ($HSI \rightarrow DJI$). The lower chart demonstrates the original closing prices of the two time series. Periods of financial stress with a significant decrease in the index prices are defined between a pair of dotted lines of the same color: the green lines represent the beginning and end of the Asian financial crisis, the red lines represent the beginning and end of the dot-com crisis, and the black lines represent the beginning and end of the 2008 global financial crisis. }\label{fig:te}\end{figure}

  The upper chart of Figure \ref{fig:te} illustrates the 30 day moving average of $TE_{DJI\rightarrow HSI}$ and $TE_{HSI\rightarrow DJI}$, as measured by  \textit{C-NJEE}. As expected, the information flow from DJI to HSI exceeds that of the opposite direction. Compered to the real prices in the lower chart of Figure \ref{fig:te}, a relatively sharp increase in $TE_{DJI\rightarrow HSI}$ is observed in times of financial stress where prices  decreasing sharply, such as in the Asian financial crisis (1997-1998), the dot-com crisis (2000-2002) and the 2008-2009 financial turmoil \cite{mcaleer2016profiteering}. This phenomenon  is well known in the financial literature (e.g., \cite{dimpfl2014impact}).

Comparing the results of the proposed method to those reported in  \cite{jiao2013universal} and \cite{zhang2019itene}, we observe that these methods also found that the information flow from DJI to HSI is much larger then in the opposite direction. However, they did not clearly determine a connection between information values and the world's financial timeline. 

\section{Conclusions}
\label{conclusion}
In this work, we introduce a neural joint entropy estimator (\textit{NJEE}). The proposed estimator is based on minimizing the CE using neural networks. Expending earlier works, we show that \textit{NJEE} is strongly consistent and provide a simple algorithmic implementation. We apply the proposed approach to  entropy estimation of random variables, specifically those with a large  alphabet,  using a simple binary transformation. Further, we introduce the conditional neural joint entropy estimator (\textit{C-NJEE}), which is an estimator for conditional joint entropy. We use \textit{NJEE} and \textit{C-NJEE} to estimate both mutual information (MI) and  conditional mutual information (CMI). 

We demonstrate the performance of the proposed schemes in  synthetic and real-world experiments.  \textit{NJEE} achieves a lower RMSE on various simulated setups
of random variables with large alphabets and  relatively small sample size. 
Moreover, the proposed MI estimator exhibits  lower bias and variance compared to newly-proposed variational lower bounds methods. This result is specifically evident in large MI values. The CMI estimator is further used to execute conditional independence tests. Again, the proposed estimator yields larger AUC value than other existing methods. Finally, we demonstrate the abilities of \textit{C-NJEE} in estimating the TE. We investigate the dynamics of information flow among financial time series and show their correlation with significant economic events. Certain important characteristics of these dynamics are not captured by other estimation methods that were implemented on the same dataset. 

We believe that future research will use the proposed entropy estimators to develop advanced compression schemes for various types of datasets. Additionally, the MI and CMI estimation capabilities can be used to improve the understanding of complex systems and deep learning frameworks. 


\section*{Acknowledgment}
The authors would like to thank Digital Living 2030 grant and the Koret foundation grant for Smart Cities and Digital Living.

\ifCLASSOPTIONcaptionsoff
  \newpage
\fi



%\bibliographystyle{IEEEtran}
\bibliography{ref}

\begin{thebibliography}{10}
\providecommand{\url}[1]{#1}
\csname url@samestyle\endcsname
\providecommand{\newblock}{\relax}
\providecommand{\bibinfo}[2]{#2}
\providecommand{\BIBentrySTDinterwordspacing}{\spaceskip=0pt\relax}
\providecommand{\BIBentryALTinterwordstretchfactor}{4}
\providecommand{\BIBentryALTinterwordspacing}{\spaceskip=\fontdimen2\font plus
\BIBentryALTinterwordstretchfactor\fontdimen3\font minus
  \fontdimen4\font\relax}
\providecommand{\BIBforeignlanguage}[2]{{%
\expandafter\ifx\csname l@#1\endcsname\relax
\typeout{** WARNING: IEEEtran.bst: No hyphenation pattern has been}%
\typeout{** loaded for the language `#1'. Using the pattern for}%
\typeout{** the default language instead.}%
\else
\language=\csname l@#1\endcsname
\fi
#2}}
\providecommand{\BIBdecl}{\relax}
\BIBdecl

\bibitem{cover2012elements}
T.~M. Cover and J.~A. Thomas, \emph{Elements of information theory}.\hskip 1em
  plus 0.5em minus 0.4em\relax John Wiley \& Sons, 2012.

\bibitem{fleuret2004fast}
F.~Fleuret, ``Fast binary feature selection with conditional mutual
  information,'' \emph{Journal of Machine learning research}, vol.~5, no. Nov,
  pp. 1531--1555, 2004.

\bibitem{peng2005feature}
H.~Peng, F.~Long, and C.~Ding, ``Feature selection based on mutual information:
  criteria of max-dependency, max-relevance, and min-redundancy,'' \emph{IEEE
  Transactions on Pattern Analysis \& Machine Intelligence}, no.~8, pp.
  1226--1238, 2005.

\bibitem{chen2016infogan}
X.~Chen, Y.~Duan, R.~Houthooft, J.~Schulman, I.~Sutskever, and P.~Abbeel,
  ``Infogan: Interpretable representation learning by information maximizing
  generative adversarial nets,'' in \emph{Advances in neural information
  processing systems}, 2016, pp. 2172--2180.

\bibitem{oord2018representation}
A.~v.~d. Oord, Y.~Li, and O.~Vinyals, ``Representation learning with
  contrastive predictive coding,'' \emph{arXiv preprint arXiv:1807.03748},
  2018.

\bibitem{tishby2000information}
N.~Tishby, F.~C. Pereira, and W.~Bialek, ``The information bottleneck method,''
  \emph{arXiv preprint physics/0004057}, 2000.

\bibitem{tishby2015deep}
N.~Tishby and N.~Zaslavsky, ``Deep learning and the information bottleneck
  principle,'' in \emph{2015 IEEE Information Theory Workshop (ITW)}.\hskip 1em
  plus 0.5em minus 0.4em\relax IEEE, 2015, pp. 1--5.

\bibitem{paninski2003estimation}
L.~Paninski, ``Estimation of entropy and mutual information,'' \emph{Neural
  computation}, vol.~15, no.~6, pp. 1191--1253, 2003.

\bibitem{wu2016minimax}
Y.~Wu and P.~Yang, ``Minimax rates of entropy estimation on large alphabets via
  best polynomial approximation,'' \emph{IEEE Transactions on Information
  Theory}, vol.~62, no.~6, pp. 3702--3720, 2016.

\bibitem{chao2003nonparametric}
A.~Chao and T.-J. Shen, ``Nonparametric estimation of shannon’s index of
  diversity when there are unseen species in sample,'' \emph{Environmental and
  ecological statistics}, vol.~10, no.~4, pp. 429--443, 2003.

\bibitem{mcallester2020formal}
D.~McAllester and K.~Stratos, ``Formal limitations on the measurement of mutual
  information,'' in \emph{International Conference on Artificial Intelligence
  and Statistics}, 2020, pp. 875--884.

\bibitem{painsky2018universality}
A.~Painsky and G.~Wornell, ``On the universality of the logistic loss
  function,'' in \emph{2018 IEEE International Symposium on Information Theory
  (ISIT)}.\hskip 1em plus 0.5em minus 0.4em\relax IEEE, 2018, pp. 936--940.

\bibitem{painsky2019bregman}
A.~Painsky and G.~W. Wornell, ``Bregman divergence bounds and universality
  properties of the logarithmic loss,'' \emph{IEEE Transactions on Information
  Theory}, vol.~66, no.~3, pp. 1658--1673, 2019.

\bibitem{hornik1989multilayer}
K.~Hornik, M.~Stinchcombe, H.~White \emph{et~al.}, ``Multilayer feedforward
  networks are universal approximators.'' \emph{Neural networks}, vol.~2,
  no.~5, pp. 359--366, 1989.

\bibitem{zhang2016understanding}
C.~Zhang, S.~Bengio, M.~Hardt, B.~Recht, and O.~Vinyals, ``Understanding deep
  learning requires rethinking generalization,'' \emph{arXiv preprint
  arXiv:1611.03530}, 2016.

\bibitem{chong2020closer}
K.~F.~E. Chong, ``A closer look at the approximation capabilities of neural
  networks,'' \emph{arXiv preprint arXiv:2002.06505}, 2020.

\bibitem{kraskov2004estimating}
A.~Kraskov, H.~St{\"o}gbauer, and P.~Grassberger, ``Estimating mutual
  information,'' \emph{Physical review E}, vol.~69, no.~6, p. 066138, 2004.

\bibitem{belghazi2018mine}
M.~I. Belghazi, A.~Baratin, S.~Rajeswar, S.~Ozair, Y.~Bengio, A.~Courville, and
  R.~D. Hjelm, ``Mine: mutual information neural estimation,'' \emph{arXiv
  preprint arXiv:1801.04062}, 2018.

\bibitem{poole2019variational}
B.~Poole, S.~Ozair, A.~v.~d. Oord, A.~A. Alemi, and G.~Tucker, ``On variational
  bounds of mutual information,'' \emph{arXiv preprint arXiv:1905.06922}, 2019.

\bibitem{song2019understanding}
J.~Song and S.~Ermon, ``Understanding the limitations of variational mutual
  information estimators,'' \emph{arXiv preprint arXiv:1910.06222}, 2019.

\bibitem{jiao2015minimax}
J.~Jiao, K.~Venkat, Y.~Han, and T.~Weissman, ``Minimax estimation of
  functionals of discrete distributions,'' \emph{IEEE Transactions on
  Information Theory}, vol.~61, no.~5, pp. 2835--2885, 2015.

\bibitem{vicente2011transfer}
R.~Vicente, M.~Wibral, M.~Lindner, and G.~Pipa, ``Transfer entropy—a
  model-free measure of effective connectivity for the neurosciences,''
  \emph{Journal of computational neuroscience}, vol.~30, no.~1, pp. 45--67,
  2011.

\bibitem{wollstadt2014efficient}
P.~Wollstadt, M.~Martinez-Zarzuela, R.~Vicente, F.~J. Diaz-Pernas, and
  M.~Wibral, ``Efficient transfer entropy analysis of non-stationary neural
  time series,'' \emph{PloS one}, vol.~9, no.~7, 2014.

\bibitem{marschinski2002analysing}
R.~Marschinski and H.~Kantz, ``Analysing the information flow between financial
  time series,'' \emph{The European Physical Journal B-Condensed Matter and
  Complex Systems}, vol.~30, no.~2, pp. 275--281, 2002.

\bibitem{dimpfl2014impact}
T.~Dimpfl and F.~J. Peter, ``The impact of the financial crisis on
  transatlantic information flows: An intraday analysis,'' \emph{Journal of
  International Financial Markets, Institutions and Money}, vol.~31, pp. 1--13,
  2014.

\bibitem{barnett2009granger}
L.~Barnett, A.~B. Barrett, and A.~K. Seth, ``Granger causality and transfer
  entropy are equivalent for gaussian variables,'' \emph{Physical review
  letters}, vol. 103, no.~23, p. 238701, 2009.

\bibitem{duan2013direct}
P.~Duan, F.~Yang, T.~Chen, and S.~L. Shah, ``Direct causality detection via the
  transfer entropy approach,'' \emph{IEEE transactions on control systems
  technology}, vol.~21, no.~6, pp. 2052--2066, 2013.

\bibitem{verdu2019empirical}
S.~Verd{\'u}, ``Empirical estimation of information measures: A literature
  guide,'' \emph{Entropy}, vol.~21, no.~8, p. 720, 2019.

\bibitem{bossomaier2016introduction}
T.~Bossomaier, L.~Barnett, M.~Harr{\'e}, and J.~T. Lizier, ``An introduction to
  transfer entropy,'' \emph{Cham: Springer International Publishing}, pp.
  65--95, 2016.

\bibitem{miller1955note}
G.~Miller, ``Note on the bias of information estimates,'' \emph{Information
  theory in psychology: Problems and methods}, 1955.

\bibitem{gao2015efficient}
S.~Gao, G.~Ver~Steeg, and A.~Galstyan, ``Efficient estimation of mutual
  information for strongly dependent variables,'' in \emph{Artificial
  intelligence and statistics}, 2015, pp. 277--286.

\bibitem{qin2019rethinking}
Z.~Qin and D.~Kim, ``Rethinking softmax with cross-entropy: Neural network
  classifier as mutual information estimator,'' \emph{arXiv preprint
  arXiv:1911.10688}, 2019.

\bibitem{mukherjee2019ccmi}
S.~Mukherjee, H.~Asnani, and S.~Kannan, ``Ccmi: Classifier based conditional
  mutual information estimation,'' \emph{arXiv preprint arXiv:1906.01824},
  2019.

\bibitem{runge2012escaping}
J.~Runge, J.~Heitzig, V.~Petoukhov, and J.~Kurths, ``Escaping the curse of
  dimensionality in estimating multivariate transfer entropy,'' \emph{Physical
  review letters}, vol. 108, no.~25, p. 258701, 2012.

\bibitem{montalto2014mute}
A.~Montalto, L.~Faes, and D.~Marinazzo, ``Mute: a matlab toolbox to compare
  established and novel estimators of the multivariate transfer entropy,''
  \emph{PloS one}, vol.~9, no.~10, p. e109462, 2014.

\bibitem{zhang2019itene}
J.~Zhang, O.~Simeone, Z.~Cvetkovic, E.~Abela, and M.~Richardson, ``Itene:
  Intrinsic transfer entropy neural estimator,'' \emph{arXiv preprint
  arXiv:1912.07277}, 2019.

\bibitem{jiao2013universal}
J.~Jiao, H.~H. Permuter, L.~Zhao, Y.-H. Kim, and T.~Weissman, ``Universal
  estimation of directed information,'' \emph{IEEE Transactions on Information
  Theory}, vol.~59, no.~10, pp. 6220--6242, 2013.

\bibitem{shalev2019context}
Y.~Shalev and I.~Ben-Gal, ``Context based predictive information,''
  \emph{Entropy}, vol.~21, no.~7, p. 645, 2019.

\bibitem{willems1998context}
F.~M. Willems, ``The context-tree weighting method: Extensions,'' \emph{IEEE
  Transactions on Information Theory}, vol.~44, no.~2, pp. 792--798, 1998.

\bibitem{liu2012relationship}
Y.~Liu and S.~Aviyente, ``The relationship between transfer entropy and
  directed information,'' in \emph{2012 IEEE Statistical Signal Processing
  Workshop (SSP)}.\hskip 1em plus 0.5em minus 0.4em\relax IEEE, 2012, pp.
  73--76.

\bibitem{zhang2018generalized}
Z.~Zhang and M.~Sabuncu, ``Generalized cross entropy loss for training deep
  neural networks with noisy labels,'' in \emph{Advances in neural information
  processing systems}, 2018, pp. 8778--8788.

\bibitem{newey1994large}
K.~Newey and D.~McFadden, ``Large sample estimation and hypothesis,''
  \emph{Handbook of Econometrics, IV, Edited by RF Engle and DL McFadden}, pp.
  2112--2245, 1994.

\bibitem{uria2016neural}
B.~Uria, M.-A. C{\^o}t{\'e}, K.~Gregor, I.~Murray, and H.~Larochelle, ``Neural
  autoregressive distribution estimation,'' \emph{The Journal of Machine
  Learning Research}, vol.~17, no.~1, pp. 7184--7220, 2016.

\bibitem{kingma2014adam}
D.~P. Kingma and J.~Ba, ``Adam: A method for stochastic optimization,''
  \emph{arXiv preprint arXiv:1412.6980}, 2014.

\bibitem{JMLR:v10:hausser09a}
\BIBentryALTinterwordspacing
J.~Hausser and K.~Strimmer, ``Entropy inference and the james-stein estimator,
  with application to nonlinear gene association networks,'' \emph{Journal of
  Machine Learning Research}, vol.~10, no.~50, pp. 1469--1484, 2009. [Online].
  Available: \url{http://jmlr.org/papers/v10/hausser09a.html}
\BIBentrySTDinterwordspacing

\bibitem{hlavavckova2007causality}
K.~Hlav{\'a}{\v{c}}kov{\'a}-Schindler, M.~Palu{\v{s}}, M.~Vejmelka, and
  J.~Bhattacharya, ``Causality detection based on information-theoretic
  approaches in time series analysis,'' \emph{Physics Reports}, vol. 441,
  no.~1, pp. 1--46, 2007.

\bibitem{dimpfl2013using}
T.~Dimpfl and F.~J. Peter, ``Using transfer entropy to measure information
  flows between financial markets,'' \emph{Studies in Nonlinear Dynamics and
  Econometrics}, vol.~17, no.~1, pp. 85--102, 2013.

\bibitem{nguyen2010estimating}
X.~Nguyen, M.~J. Wainwright, and M.~I. Jordan, ``Estimating divergence
  functionals and the likelihood ratio by convex risk minimization,''
  \emph{IEEE Transactions on Information Theory}, vol.~56, no.~11, pp.
  5847--5861, 2010.

\bibitem{campos2006scoring}
L.~M.~d. Campos, ``A scoring function for learning bayesian networks based on
  mutual information and conditional independence tests,'' \emph{Journal of
  Machine Learning Research}, vol.~7, no. Oct, pp. 2149--2187, 2006.

\bibitem{sen2017model}
R.~Sen, A.~T. Suresh, K.~Shanmugam, A.~G. Dimakis, and S.~Shakkottai,
  ``Model-powered conditional independence test,'' in \emph{Advances in Neural
  Information Processing Systems}, 2017, pp. 2951--2961.

\bibitem{zhang2012kernel}
K.~Zhang, J.~Peters, D.~Janzing, and B.~Sch{\"o}lkopf, ``Kernel-based
  conditional independence test and application in causal discovery,''
  \emph{arXiv preprint arXiv:1202.3775}, 2012.

\bibitem{sachs2005causal}
K.~Sachs, O.~Perez, D.~Pe'er, D.~A. Lauffenburger, and G.~P. Nolan, ``Causal
  protein-signaling networks derived from multiparameter single-cell data,''
  \emph{Science}, vol. 308, no. 5721, pp. 523--529, 2005.

\bibitem{ben2008bayesian}
I.~Ben-Gal, ``Bayesian networks,'' \emph{Encyclopedia of statistics in quality
  and reliability}, vol.~1, 2008.

\bibitem{statnikov2013algorithms}
A.~Statnikov, N.~I. Lytkin, J.~Lemeire, and C.~F. Aliferis, ``Algorithms for
  discovery of multiple markov boundaries,'' \emph{Journal of Machine Learning
  Research}, vol.~14, no. Feb, pp. 499--566, 2013.

\bibitem{gers1999learning}
F.~A. Gers, J.~Schmidhuber, and F.~Cummins, ``Learning to forget: Continual
  prediction with lstm,'' 1999.

\bibitem{mcaleer2016profiteering}
M.~McAleer, J.~Suen, and W.~K. Wong, ``Profiteering from the dot-com bubble,
  subprime crisis and asian financial crisis,'' \emph{The Japanese Economic
  Review}, vol.~67, no.~3, pp. 257--279, 2016.

\end{thebibliography}
\bibliographystyle{IEEEtran}


% \begin{IEEEbiography}{Yuval Shalev}
% \end{IEEEbiography}

% \begin{IEEEbiographynophoto}{Amichai Painsky}

% % insert where needed to balance the two columns on the last page with
% %\newpage

% \begin{IEEEbiographynophoto}{Irad Ben-Gal}


%\vfill


\end{document}